\documentclass{jfm}
\usepackage{graphicx,bm,amsmath}
\usepackage{epstopdf, epsfig}
\usepackage[usenames,dvipsnames,svgnames,table]{xcolor}
\usepackage{caption,subcaption}

\newcommand{\Dpol}{\pi_{\Delta y}}
\shorttitle{Synchronised Swimming of Two Fish}
\shortauthor{Novati,  Verma, Alexeev, Rossinelli, van Rees and Koumoutsakos}

\title{Synchronised Swimming of Two Fish}

\author{Guido Novati\aff{1}, Siddhartha Verma\aff{1}, Dmitry Alexeev\aff{1},  Diego Rossinelli\aff{1}, Wim M. van Rees\aff{1,2} \and Petros Koumoutsakos\aff{1}\corresp{\email{petros@ethz.ch}}}

\affiliation{\aff{1}Computational Science and Engineering Laboratory, \\ Clausiusstrasse 33, ETH Z\"{u}rich, CH-8092, Switzerland
\aff{2}School of Engineering and Applied Sciences, Harvard University,USA}

\DeclareMathOperator*{\argmax}{arg\,max}

\begin{document}

\maketitle

\begin{abstract}
We study the fluid dynamics of two fish-like bodies with synchronised swimming patterns. Our studies are based on two-dimensional simulations of viscous incompressible flows. We distinguish between motion patterns that are externally imposed on the swimmers and self-propelled  swimmers that learn manoeuvres to achieve certain goals. Simulations of two rigid bodies executing pre-specified motion indicate that flow-mediated interactions can lead to substantial drag reduction and may even generate thrust intermittently. In turn we examine  two self-propelled swimmers arranged in a leader-follower configuration, with a-priori specified body-deformations. We find that the swimming of the leader remains largely unaffected, while  the follower experiences either an increase or decrease in swimming speed, depending on the initial conditions. Finally, we consider a follower that synchronises its motion so as to minimize its lateral deviations from the leader's path. The leader employs a steady gait  while the follower uses a reinforcement learning algorithm to adapt its swimming-kinematics. We find that swimming in a synchronised tandem can yield up to about $30\%$ reduction in energy expenditure for the follower, in addition to a $20\%$ increase in its swimming-efficiency. The present results indicate that synchronised swimming of two fish can be energetically beneficial.
\end{abstract}


\section{Introduction}

The coordinated motion of fish is thought to provide an energetic advantage to individuals, as well as to the group as a whole, in terms of increased swimming range, endurance and chances of survival. Schooling has been credited with serving various other biological functions including defence from predators~\citep{Brock1960,Cushing1968}, enhanced feeding and reproductive opportunities~\citep{Shaw1978,Pitcher1982}. Amid a lack of clear agreement regarding the evolutionary purpose of schooling behaviour~\citep{Pavlov2000}, there is growing evidence supporting the role of hydrodynamic mediation in facilitating propulsion. Experiments investigating fish swimming in groups point to a reduction in energy expenditure, based on respirometer readings and reduced tail-beat frequency~\citep{Parker1973,Abrahams1985,Herskin1998,Svendsen2003,Killen2012}. 
Importantly, there is evidence to suggest that reduction in energy expenditure is not distributed uniformly throughout a schooling group. \citet{Herskin1998,Svendsen2003,Killen2012} observed that the tail-beat frequency of trailing fish was lower than that of fish at the front of the school. Moreover, \citet{Killen2012} note that fish with inherently lower aerobic scope  prefer to stay towards the rear of a group. Studies investigating the response of solitary fish to unsteady flow~\citep{Liao2003Science} found that trout swimming behind obstacles  exerted reduced effort for station-keeping. The trout adopted a modified gait, which allowed them to `slalom' through the oncoming vortex street. The ensuing reduction in muscle activity was confirmed using neuromuscular measurements~\citep{Liao2003Science} and respirometer readings~\citep{Taguchi2011}. These experimental studies suggest that fish can detect and exploit vortices  present in their surroundings.

There is a well documented hypothesis~\citep{Breder1965,Weihs1973,Weihs1975,Shaw1978} that flow patterns which emerge as a consequence of schooling, can be exploited by individual swimmers. This hypothesis was first quantified~\citep{Weihs1973,Weihs1975} by using inviscid point-vortices as models of the fish wake-vortices. It was postulated that large groups of fish could gain a propulsive advantage by swimming in a `diamond' configuration, with opposing tail-beat phase. The energetic gain was attributed to two distinct mechanisms: drag reduction resulting from decreased relative velocity in the vicinity of specific vortices; and a forward `push' originating from a `channelling effect' between lateral neighbors.  Weihs noted that a rigid geometrical arrangement, and perfectly synchronized anti-phase swimming among lateral neighbours, were unlikely to occur in nature. Nonetheless, he postulated that given the immense potential for energy savings, even intermittent utilization of the proposed arrangement could lead to a tangible benefit~\citep{Weihs1975}.  The role of hydrodynamics in fish-schooling was later questioned~\citep{Partridge1979}, based on empirical observations of fish-schools which rarely displayed diamond formations. However, a later study based on aerial photographs of hunting tuna schools~\citep{Partridge1983} provided evidence for such diamond-like formations. We believe that these studies highlight the difficulties of maintaining fixed patterns in the dynamically evolving environment of schooling fish. 
These difficulties are also reflected in simulations studies. Theoretical and numerical studies of schooling have employed potential flow models~\citep{Tsang2013,Gazzola2016}, or they have pre-specified the spatial distribution and motion of the swimmers~\citep{Hemelrijk2015,Daghooghi2015}. 

Here we present two-dimensional simulations of viscous, incompressible flows of multiple self-propelled swimmers that can dynamically adapt their motion. We focus in particular on two swimmers arranged in a  leader-follower configuration. We investigate the hydrodynamic interactions of the swimmers in various scenarios including pre-specified coordinated motions and initial distances, as well as the dynamic adaptation of the follower's motion using a reinforcement learning algorithm, so as to remain within a specific region in the leader's wake.  We investigate the impact of the leader's wake on the follower's motion and identify the mechanisms that lead to energy savings. The paper is organised as follows: we outline the numerical methods for the simulations of self-propelled swimmers in Section~\ref{sec:simDetails}, and the  reinforcement learning algorithm is discussed in Section~\ref{sec:learning}. We present results of the three synchronised swimming scenarios in Section~\ref{sec:Results}, followed by concluding remarks in Section~\ref{sec:Conclusion}.

\section{Simulation details}
\label{sec:simDetails}

\subsection{Swimmers and Numerical Methods}
\label{sec:NumMeth}

We solve the  two-dimensional, viscous, incompressible Navier-Stokes equations in velocity-vorticity form using remeshed vortex methods on wavelet-adapted grids~\citep{Rossinelli2015}, and a divergence-free penalisation technique to enforce the no-slip boundary condition~\citep{Gazzola2011}. The wavelet adaptivity and the computational efficiency of the solver are critical aspects for this work as they enable the utilisation of the costly reinforcement learning algorithms.
 
The self-propelled swimmers used in the simulations are based on a simplified physical model of zebrafish as described in~\citet{Gazzola2011}. 
 Undulations of the swimmer's body are generated by imposing a spatially and temporally varying body curvature ($k(s,t)$), which passes down from the head to the tail as a sinusoidal travelling wave:
\begin{equation}
k(s,t) = A(s) \ \sin{\left[ 2 \pi \left( \frac{t}{T_{p}} - \frac{s}{L} \right) + \phi\right]}  
\label{eq_ks}
\end{equation}
Here, $L$ is the length of the swimmer, $T_p = 1$ is the tail-beat frequency, $\phi$ is the phase-difference. The  curvature amplitude $A(s)$ varies linearly from the head to the tail thus reducing head motion and amplifying  tail-beat amplitude.
We estimate the swimming-efficiency using a modified form of the Froude efficiency proposed by~\citet{Tytell2004}: 
\begin{equation}
\eta = \dfrac{P_{thrust}}{P_{thrust} + \max\left(P_{def},\ 0\right)} = \dfrac{T u}{T u  + \max\left(-\iint \bm{u}_{def} \cdot \bm{dF}, \ 0\right)}
\label{eq:eta}
\end{equation}
$P_{thrust}$ and $P_{def}$ represent the power output related to thrust generated by the body, and the power exerted in deforming the swimmer's body against fluid-induced forces. The $\max$ operator effectively clips the maximum of $\eta$ to 1. This is necessary to avoid undefined values of $\eta$, which can occur when fluid-induced surface-forces and the deformational velocity ($\bm{u}_{def}$) point in the same direction, giving rise to negative $P_{def}$. The thrust is computed as $T = \iint(\bm{u}\cdot\bm{dF} + \lvert \bm{u}\cdot\bm{dF}\rvert)/2$, where $\bm{dF}=\bm{dF}_P + \bm{dF}_\nu$ is comprised of the viscous- and pressure-based forces acting on the swimmer:
\begin{equation}
\bm{dF}_\nu = 2\mu \bm{D}\cdot \bm{n} \ dS \quad \text{and} \quad \bm{dF}_P = -P \bm{n} \ dS
\label{eq:fPress}
\end{equation}
Here, $\bm{D} = (\nabla\bm{u} + \nabla\bm{u}^T)/2$ is the strain-rate tensor, $P$ is the surface-pressure, and $\mu$ is the dynamic viscosity, $\bm{n}$ represents the surface-normal and $dS$ denotes the corresponding infinitesimal surface area. The surface-pressure is obtained by solving a  Poisson's equation ($ \nabla^2 P = -\rho\left(\nabla\bm{u}^T:\nabla\bm{u}\right) + \rho\lambda\nabla\cdot\left(\chi\left(\bm{u}_s-\bm{u}\right)\right)$) \citep{Verma2016}.
 
\subsection{Reinforcement learning}
\label{sec:learning}
The two swimmers either follow a-priori defined swimming patterns, or the follower adapts its body-deformation to synchronise its motion with that of the leader. This adaptation is achieved using a Reinforcement Learning (RL) \citep{Sutton1998}, a potent machine learning algorithm for model-free flow control ~\citep{Gautier2015,Gazzola2016}. In RL, the agents   receive information about their {\it State} and chose {\it Actions} to maximise a cumulative future {\it Reward} in an unsupervised manner. 
The swimmer learns to estimate the action-value $Q(s,a)$ which is defined as the expected sum of the discounted future rewards $r(s,a,s')$, for each action $a$ being performed in each state $s$. The reward is obtained by starting in $s$, performing $a$ to end up in a new state $s'$, and thereafter following the policy $\pi(s')$:
\begin{equation}
Q_\pi (s,a) = \mathbb{E} \left[ r(s,a,s') + \gamma Q(s',\pi(s')) \right]
\label{eq:qPi}
\end{equation}
The discount factor $\gamma \in [0,1] $ determines the trade-off between immediate and future rewards and was set to $\gamma=0.8$ for all present results. The learning process terminates upon convergence of $Q(s,a)$, and the swimmer can make optimal decisions by following a `greedy' policy ($\pi(s) = \argmax_a Q_\pi (s,a)$).
The result of this process is usually a tabular approximation of state-action-reward values.
An important note for the present study is the realisation that the motion of the swimmer implies a continuous state space. Hence, in this work the swimmer learns a parametrized approximation of $Q(s,a)$ by training a Neural Network (NN) with experience replay~\citep{Lin1992}. The algorithm involves storing all observed transitions $\{s,a,s',r\}$, which are sampled to update the value function.

\begin{figure}
  \centering
  \begin{subfigure}[b]{0.66\textwidth}
    \centering
    \includegraphics[]{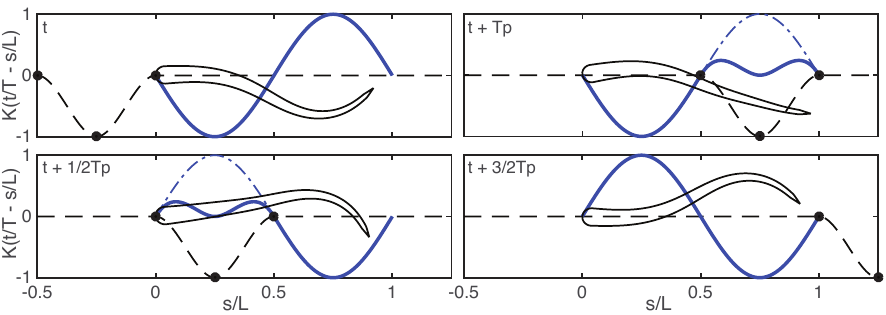}
    \subcaption{}
    \label{fig:fig_spline}
  \end{subfigure}
  \begin{subfigure}[b]{0.32\textwidth}
    \centering
    \includegraphics[]{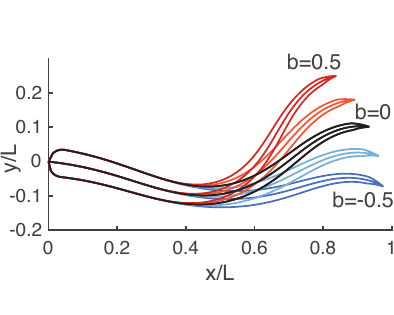}
    \subcaption{}
    \label{fig:fig_act} 
  \end{subfigure}
  \caption{(\subref{fig:fig_spline}) Modification of the swimmer's midline curvature (solid blue line) by superimposing opposing curvature with $b=-1$. The black dashed line corresponds to $M(t/T_p - s/L)$ in Eq.~\ref{eq:turn}, and the blue dash-dot line indicates the unmodified curvature (i.e., the sinusoidal part of Eq.~\ref{eq_ks}). (\subref{fig:fig_act}) The impact of varying the control-amplitude on the modified shape. The unmodified shape corresponds to $b=0$.}
\end{figure}

{\it Actions} taken by a swimmer involve manipulating its body curvature in a manner which allows it to execute turns and  to control its speed. This is achieved by introducing a linear superposition to the travelling wave described in Eq.~\ref{eq_ks}:
\begin{equation}
k_{learner}(s,t) = k(s,t) + k'(s,t) = A(s) \left( sin{\left[ 2 \pi \left( \frac{t}{T_{p}} - \frac{s}{L} \right) + \phi \right]}  + M\left( \frac{t}{T_{p}} - \frac{s}{L} \right) \right)  
\label{eq:turn}
\end{equation}
where $M\left( \frac{t}{T_{p}} - \frac{s}{L} \right)$ defines a travelling natural cubic spline, computed using three evenly-spaced nodes separated by a distance of $L/4$ (figure~\ref{fig:fig_spline}). The spline shape  remains constant as it proceeds from the head to the tail, and is determined by a fixed control-amplitude `$b$' associated with a particular action. The swimmers are permitted to execute an action every half tail-beat period ($T_p/2$), which allows them to either increase or reduce the undulation amplitude (figures~\ref{fig:fig_spline} and~\ref{fig:fig_act}).

\begin{figure}
  \centering
  \includegraphics{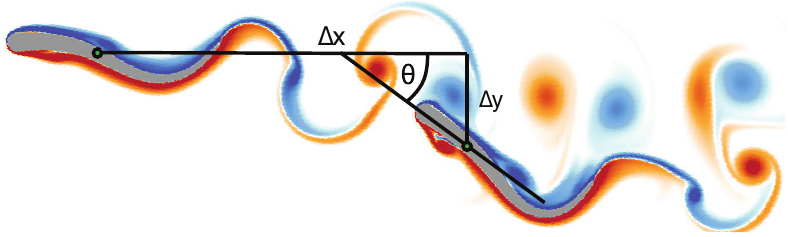}
	\caption{The leader swims along the horizontal line, the follower perceives its displacement and inclination relative to the leader.}
  \label{fig:fig_prob}
\end{figure}

In the present two fish swimming (leader-follower tandem) problem,  learning is performed only by the follower.
The  follower defines its current {\it State} using its displacement ($\Delta x$ and $\Delta y$) and orientation ($\theta$) relative to the leader (figure~\ref{fig:fig_prob}).
Moreover, the follower also considers whether it is taking an action in the first or second half of its tail-beat period, as part of its state ($\text{modulo}(\textrm{time}, T_p)/T_p$). This is necessary because the same action can either increase or decrease the mid-line curvature, depending on whether it is taken at $T_P$ or at $T_P/2$ (figure~\ref{fig:fig_spline}). Furthermore, the swimmer performs two actions every period, as described earlier. The effect of each action travels along the swimmer's body, affecting its interaction with the flow over the next swimming period. For this reason, the state of the swimmer also includes the two actions taken over the previous tail-beat period.

Here, the {\it Reward} used to provide feedback to the  follower regarding its performance is defined as:
$
\mathcal{R}_{\Delta y} = 1 - \dfrac{| \Delta y | }{0.5 L}
$.
This function penalizes the follower when it laterally strays too far from the path of the leader.
The follower learns the policy $\pi_{\Delta y}$ which maximizes $\mathcal{R}_{\Delta y}$. The state space is restricted to $1 \leq \Delta x/L \leq 3$, $ | \Delta y / L| \leq 0.5$ and $ | \theta | \leq \pi/2$. When an action leads the follower to exceed these thresholds, the learner transitions to a terminal state with reward $\mathcal{R}_{\Delta y}=-1$, and the simulation is terminated.

\section{Results}
\label{sec:Results}

We distinguish externally imposed motions on the swimmers to those that are achieved by the deformation of the body of self-propelled swimmers. Results concerning three distinct scenarios, namely, two rigid airfoils executing pre-specified motion, two self-propelled swimmers interacting without control, and a `smart' follower utilizing adaptive control to interact with the leader's wake, are discussed in this section.
\subsection{Rigid objects with pre-specified motion}
\label{sec:rigid}

A  swimming pattern often observed in schooling involves the exchange of the leader and follower positions. We examine this scenario by first studying two rigid, airfoil-shaped  bodies (shape identical to swimmers) dragged along prescribed sinusoidal paths to exchange their position as leader and follower (see Movie 1):
\begin{equation}
a_x(t) = \begin{cases}
- \frac{u_{max}-u_{min}}{T_s}   & \ \text{for } \ 2 n T_s \leq t < (2 n T_s +1) \\
  \frac{u_{max}-u_{min}}{T_s}  & \ \text{for }  \ (2 n T_s +1) \leq t < 2n (T_s +1) 
  \end{cases} 
\label{eq:Dead_accx}
\end{equation}
Here, $n$ is an integer, $u_{max}=4.5L/T_s$, $u_{min}=1.5L/T_s$, and $T_s$ represents the time-period with which the bodies exchange their position as leader and follower. The vertical displacement of the center-of-mass is determined as $y(\Delta x,L) = L/5\cos { (\pi \Delta x/L)}$, where $\Delta x$ is the horizontal distance traversed. The orientation of the airfoils is aligned with the tangents of their respective trajectories. Both the airfoils start their motion at the same $x$-location; one of the objects is initialized at a crest with $u_{max}$ and undergoes steady deceleration (Eq~\ref{eq:Dead_accx}), whereas the other object starts with $u_{min}$ on a trough and is subjected to constant acceleration. This arrangement of positions and velocities alternates between the two airfoils every time-period $T_s$.

\begin{figure}
     \centering
     \begin{subfigure}[b]{0.54\textwidth}
 		\centering
                \includegraphics[]{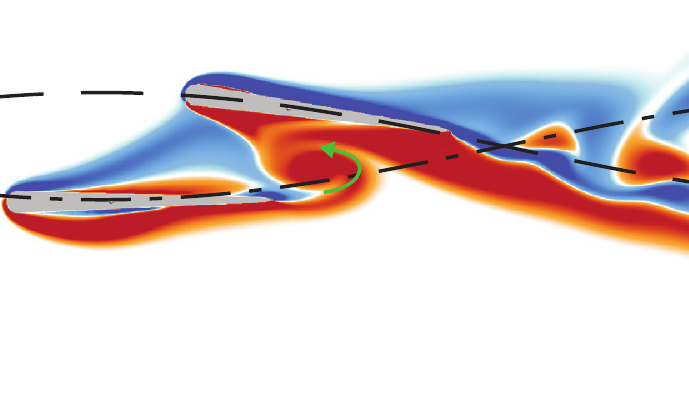}
                \subcaption{}
		\label{fig:DeadVort}
      \end{subfigure}
       \begin{subfigure}[b]{0.44\textwidth}
       		\centering
                \includegraphics[]{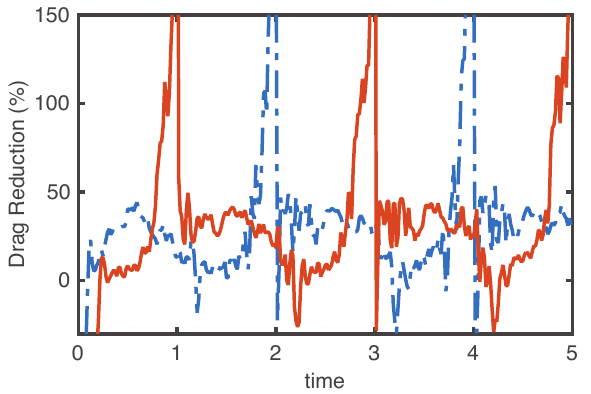}
                \subcaption{}
		\label{fig:DeadDrag}
      \end{subfigure}
	\caption{(\subref{fig:DeadVort}) Vorticity field generated by the two fish-shaped airfoils dragged with the prescribed sinusoidal pattern ($Re = u_{max} L / \nu = 2250$). The snapshot shown corresponds to $t=3.88$.
(\subref{fig:DeadDrag}) Percentage of drag reduction for the two airfoils, with respect to a single airfoil executing the same motion pattern.}
	\label{fig:deadFish}
\end{figure}

A snapshot of the resulting vorticity field, along with the sinusoidal path followed by the two airfoils, is shown in figure~\ref{fig:deadFish}.
The flow pattern that emerges influences the net drag acting on the two objects. We observe that the follower in figure~\ref{fig:DeadVort} experiences a dramatic reduction in drag, as indicated by the dash-dotted line in figure~\ref{fig:DeadDrag} ($t\approx3.9$). This can be attributed to a decrease in relative velocity, due to the presence of the positive vortex highlighted in figure~\ref{fig:DeadVort}. The drag reduction at this time instance is greater than $100\%$, which corresponds to a net thrust being generated due to the interaction of the follower's motion with the wake. Moreover, figure~\ref{fig:DeadDrag} indicates that both the leader and the follower may experience a reduction in drag as a result of mutual interaction. The results suggest that hydrodynamic interactions between solid objects executing specific motion patterns can give rise to substantial drag reduction, and even intermittent thrust production.

\subsection{Tandem of two-self propelled swimmers: No Control}
\label{sec:passive}
We examine the behaviour of a self-propelled swimmer placed initially in a tandem configuration with a leader. Both swimmers have a-priori defined body-deformations. Their motions are determined by their deformations and, in particular for the follower, by their interaction with the flow field.  The kinematics imposed for body-undulations are identical for both the leader and the follower (Eq.~\ref{eq_ks}), and correspond to a Reynolds number of $Re = L^2 / T_P \nu = 5000 $. The crucial difference from the configuration studied in Section~\ref{sec:rigid} is that both the swimmers are now self-propelled (Section~\ref{sec:NumMeth}), and their trajectories are not imposed, but emerge from their interaction with the flow.

\begin{figure}
	\centering
	\begin{subfigure}[b]{\textwidth}
       		\centering
		\begin{minipage}{.49\textwidth}
                \includegraphics{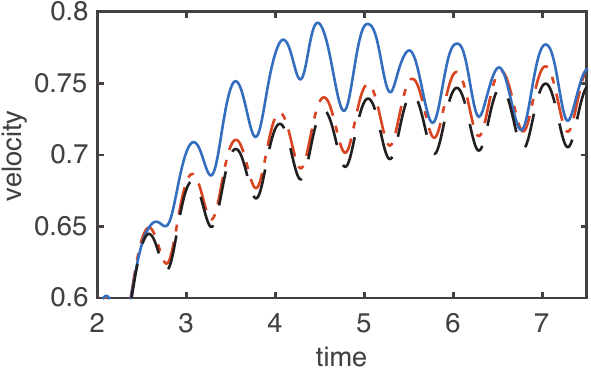}
		\end{minipage}
		\begin{minipage}{.49\textwidth}
                \includegraphics{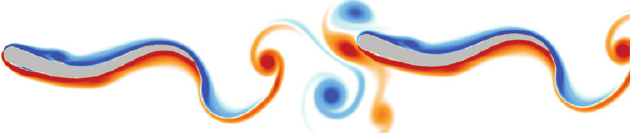}
		\vspace{0.7cm}
		\end{minipage}
		\subcaption{}
                \label{fig:PassiveGood}
     	\end{subfigure}
     	\begin{subfigure}[b]{\textwidth}
     		\centering
		\begin{minipage}{.49\textwidth}
                \includegraphics{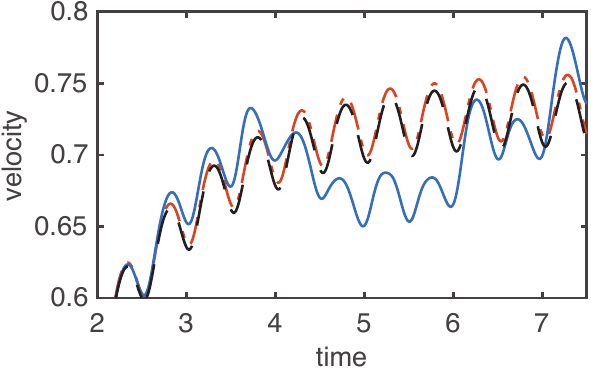}
		\end{minipage}
		\begin{minipage}{.49\textwidth}
                \includegraphics{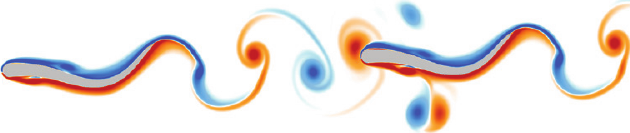}
		\vspace{0.8cm}
		\end{minipage}
                \subcaption{}
		\label{fig:PassiveBad}
      	\end{subfigure}
	\caption{The velocity of the leader (dash-dot orange line), the follower (solid blue line), and a solitary swimmer (dashed black line) for (\subref{fig:PassiveGood}) $\delta_0 = 1.75L$ (with the vorticity field shown at $t=2.6$) and (\subref{fig:PassiveBad}) $\delta_0 = 2.15L$  (vorticity field shown at $t=3.6$).}
	\label{fig:Passive}
\end{figure}

We consider two different cases, with the leader and the follower starting from rest at a separation distance of $\delta_0 = 1.75L$ with $\phi=0$ (figure~\ref{fig:PassiveGood}), and at $\delta_0 = 2.15L$ with $\phi=\pi/2$ (figure~\ref{fig:PassiveBad}).
The vorticity fields shown in both figures~\ref{fig:PassiveGood} and~\ref{fig:PassiveBad} correspond to instances when the follower first encounters the leader's wake. We observe an increase of $9.5\%$ in the follower's maximum velocity in the first case (figure~\ref{fig:PassiveGood}, $t\approx 4.4$), whereas the follower in the second case experiences a velocity reduction of up to $-9.1\%$ (figure~\ref{fig:PassiveBad}, $t\approx4.9$). 

These results suggest that hydrodynamic interaction with a leader's wake can have both a beneficial, as well as a detrimental impact on the performance of a follower. Furthermore, in both cases, the follower's trajectory starts deviating laterally as soon as it encounters the wake, and the follower is completely clear of the wake after approximately 4 to 6 tail-beat periods (supplementary Movie 2). This suggests the need for active modulation of the trailing swimmer's actions when navigating the wake of a leader, in order to maintain a tandem configuration.

\subsection{Tandem of two-self propelled swimmers: Adaptive Control}

In this section, we discuss the swimming-efficiency of a  follower that adapts its motion using a RL algorithm, in response to velocity fluctuations in the leader's wake. The steady gait of the leader corresponds to a Reynolds number of $Re = L^2 / T_P \nu = 5000 $. In order to ensure adequate exploration of the state space, the follower initially performs random actions with a $50\%$ probability, which is gradually reduced to $10\%$. The policy $\Dpol$ is determined using approximately 100,000 state-actions-reward sets.

\begin{figure}
  \centering
  \begin{subfigure}[b]{0.49\textwidth}
    \centering
    \includegraphics[]{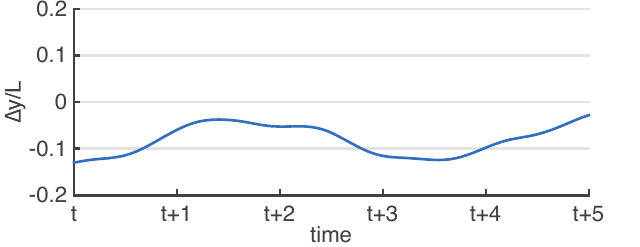}
    \subcaption{}
    \label{fig:DY}
    \end{subfigure}
  \begin{subfigure}[b]{0.49\textwidth}
    \centering
    \includegraphics[]{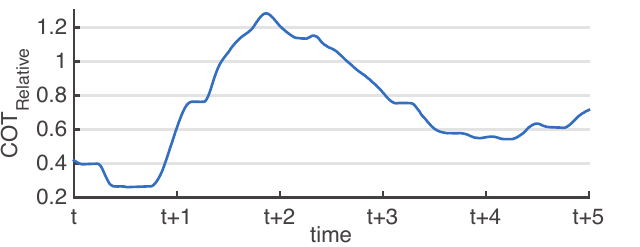}
    \subcaption{}
    \label{fig:COT}
    \end{subfigure}
  \caption{(\subref{fig:DY}) Lateral displacement, and (\subref{fig:COT}) CoT (equation~\ref{eq:COT}) of the  follower, normalized with the CoT of a solitary swimmer.}
  \label{fig:deltaYcot}
\end{figure}

The  follower actively seeks to maintain its position in the center of the leader's wake, as can be observed from the time-evolution of $\Delta y$ in figure~\ref{fig:DY} (and supplementary Movie 3). 
Without active control, the follower may get deflected away from the wake, as discussed in section~\ref{sec:passive}. To examine the impact that minimizing $\lvert\Delta y\rvert$ has on the energy consumption of the follower, we compute the `Cost of Transport' (CoT) as follows:
\begin{equation}
\textrm{CoT}(t) = \frac{\int^{t}_{t-T_p} \max(P_{def},0)  \textrm{d}t}{\int^{t}_{t-T_p} \| \bm{u} \| \textrm{d}t}  \label{eq:COT}
\end{equation}
\begin{figure}
	\centering
	\begin{subfigure}[b]{\textwidth}
		\centering
		\includegraphics[]{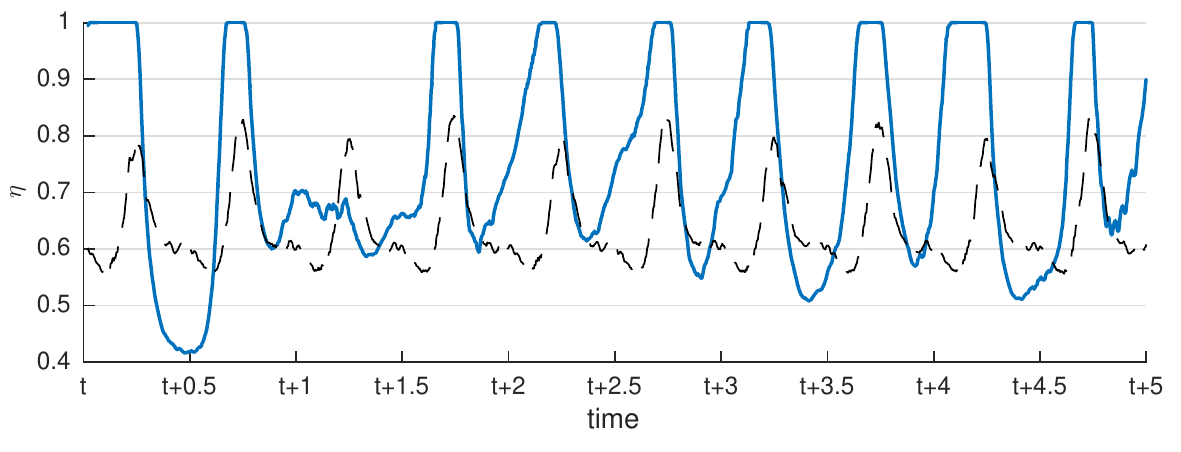}
		\subcaption{}
		\label{fig:eff3}
	\end{subfigure}
	\begin{subfigure}[b]{0.49\textwidth}
		\centering
		\includegraphics[width=\textwidth]{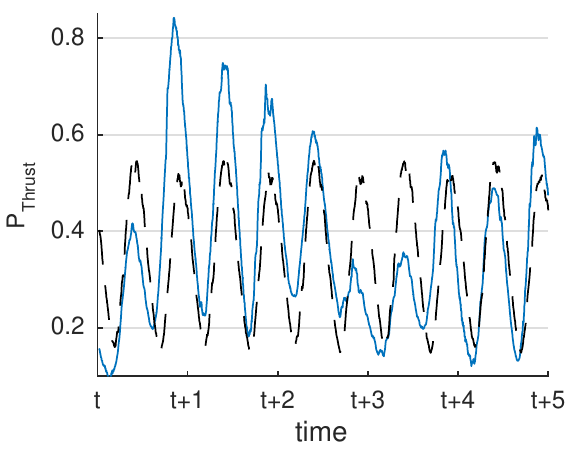}
		\subcaption{}
		\label{fig:pT3}
	\end{subfigure}
	\begin{subfigure}[b]{0.49\textwidth}
		\centering
		\includegraphics[width=\textwidth]{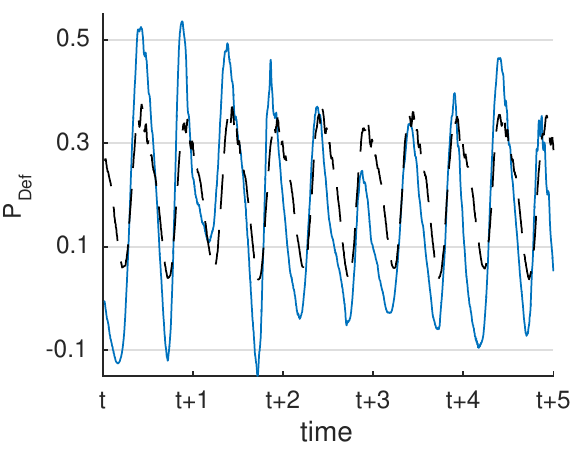}
		\subcaption{}
		\label{fig:pDef3}
	\end{subfigure}
	\caption{(\subref{fig:eff3}) Efficiency,  (\subref{fig:pT3}) $P_{thrust}$, and (\subref{fig:pDef3}) $P_{def}$ measured over 5 swimming periods for a solitary passive swimmer (dashed black line), and for the follower acting according to $\Dpol$ (solid blue line). Power measurements are non-dimensionalized by $m L^2 / T_p^3$ ($m$ represents the mass of the swimmer).}
	\label{fig:effs}
\end{figure}
The CoT indicates the energy spent per unit distance travelled. We remark that the $\max$ operator in Eq.~\ref{eq:COT} precludes  negative values of $P_{Def}$, and accounts for the fact that the trailing swimmer may not `store' energy from the flow. Figure~\ref{fig:COT} shows the CoT for the active trailing fish, normalized with respect to the CoT of an isolated swimmer (which remains approximately constant during steady-state swimming). The relative CoT tends to be smaller than $1$ for the most part, which is indicative of the  follower spending less energy in traversing a unit distance compared to an isolated swimmer. There are instances when the relative CoT exceeds $1$, which corresponds to the follower exerting additional effort to execute corrective actions. Over a course of $40$ time-periods, actively minimizing the lateral distance from the leader results in substantial energy savings per unit distance travelled; the swimmer following $\Dpol$ requires $29.3\%$ less energy than a solitary swimmer in an unperturbed flow. These results showcase the significant energetic benefits that can be obtained by a follower when exploiting a leader's wake.

In addition to the CoT, we examine the efficiency $\eta$ (Eq.~\ref{eq:eta}), the thrust-related power $P_{thrust}$, and power consumed by body-deformation $P_{Def}$ for a solitary swimmer adopting a steady gait, and the follower acting according to $\Dpol$ (figure~\ref{fig:effs}). The active follower experiences an increase in swimming-efficiency (figure~\ref{fig:eff3}). This points to an ability to extract energy from the oncoming vortices, and a consequential reduction in effort exerted by the swimmer. The net increase in average efficiency over a duration of $40$ time-periods is approximately $19.4\%$. These gains do not arise due to an increase in $P_{Thrust}$ (figure~\ref{fig:pT3}), but rather due to a reduction in $P_{Def}$ (figure~\ref{fig:pDef3}). The time-averaged $P_{Thrust}$ over $40$ time-periods varies only moderately for the active swimmer compared to the solitary swimmer ($-1.2\%$), but the reduction in $P_{Def}$ is substantial ($-36.6\%$). This may be related to the fact that the  follower tends to favour a decrease, rather than an increase, in the undulation-amplitude (section~\ref{sec:learning}).

The thrust- and deformation-power for the  follower show a noticeable variation in amplitude, compared to those for a solitary swimmer (figure~\ref{fig:pT3}). The fluctuating power-output is related to the fact that the distance-based reward $\mathcal{R}_{\Delta y}$ solely aims to minimize $\Delta y$. This can affect efficiency adversely for relatively short durations, as observed at times $(t+0.5)$, $(t+3.5)$ and $(t+4.5)$ in figure~\ref{fig:eff3}. Nonetheless, the `smart' follower is more efficient on average than a solitary swimmer ($19.4\%$ higher efficiency). This indicates that the substantial reduction in energy consumption ($29.3\%$ drop in relative CoT) does not come at the expense of decreased efficiency.

\section{Conclusion}
\label{sec:Conclusion}

In this paper, we demonstrate the energetic benefits of coordinated swimming, for two swimmers in a leader-follower configuration through a series of simulations. First, an arrangement of rigid airfoil-shaped swimmers, executing pre-specified motion, is observed to give rise to substantial drag reduction. Following this, we investigate self-propelled fish shapes, with both the leader and the follower employing identical kinematics. Without any active adaptation, the follower's interactions with the leader's wake can be either energetically beneficial or detrimental, depending on the initial condition. Furthermore, the follower tends to diverge from the leader's wake, which points to the need for active modulation of the follower's actions to maintain a stable tandem configuration. Finally, we examine, for the first time to the best of our knowledge, the case where the leader swims with a steady gait and the follower adapts its behaviour dynamically to account for the effects of the wake encountered. The actions of the follower are selected autonomously from an optimal policy determined via reinforcement learning, and allow the swimmer to maximize a specified long-term reward. The results indicate that swimming in tandem can lead to measurable energy savings for the follower. We measure about 30\% reduction in energy spent per unit distance, compared to a solitary swimmer, even when the goal of the follower is to minimize lateral distance from the leader. The results demonstrate that, for two fish swimming in a synchronised tandem configuration, can give rise to substantial energetic benefits. 

Current work extends these simulations to three-dimensional flows by deploying wavelet adapted vortex to massively parallel computing architectures. We envision that this work will be important for the design of energetically efficient robotic devices that need to account for strong hydrodynamic interactions.

\section*{Acknowldgement}
This work utilized computational resources granted by the Swiss National Supercomputing Centre (CSCS) under project ID `s436'. We gratefully acknowldge support by the European Research Council Advanced Investigator Award.

\bibliographystyle{jfm}
\bibliography{twofish}

\end{document}